\documentclass[twocolumn,showpacs,preprintnumbers,amsmath,amssymb]{revtex4}

\usepackage{epsfig}
\usepackage{graphicx}
\usepackage{dcolumn}
\usepackage{bm}

\def\DESepsf(#1 width #2){\epsfxsize=#2 \epsfbox{#1}}

\begin{document}


\title{\boldmath{Accounting for Slow $J/\psi$ from $B$ Decay}}

\author{$^{a)}$Chun-Khiang Chua}
\author{$^{a)}$Wei-Shu Hou}%
\author{$^{b)}$Gwo-Guang Wong}
\affiliation{%
$^{a)}$Department of Physics, National Taiwan University, Taipei,
Taiwan 10764, Republic of China\\
$^{b)}$Department of International Trade, Lan Yang Institute of
Technology, Toucheng, Ilan, Taiwan 26141, Republic of China
}%

\date{\today}

\begin{abstract}
A slow $J/\psi$ excess exists in the inclusive $B\to J/\psi+X$
spectrum, and is indicative of some hadronic effect. From color
octet nature of $c\bar c$ pair in $b\to c\bar cs$ decay, one such
possibility would be $B \to J/\psi+ K_g$ decay, where $K_g$ is a
hybrid resonance with $\bar sgq$ constituents. We show that a
$K_g$ resonance of $\sim$ 2 GeV mass and suitably broad width
could be behind the excess.
\end{abstract}

\maketitle

Using 1.12 fb$^{-1}$ data, the CLEO experiment published the
inclusive $B\to J/\psi+X$ spectrum~\cite{CLEO95}. After
subtracting $\chi_{c1}$, $\chi_{c2}$ and $\psi(2S)$ feeddown,
there was a hint of excess events around $p^*_{J/\psi}\sim 0.5$
GeV, where $p^*_{J/\psi}$ is $J/\psi$ momentum in $\Upsilon(4S)$
frame.
With 6.2 fb$^{-1}$ data, the Belle experiment
presented~\cite{Schrenk} the inclusive $B\to J/\psi+X$ spectrum.
After feeddown subtraction, one could also infer~\cite{IC} an
excess for $p^*_{J/\psi}\lesssim 0.8$ GeV.
Recently, the BaBar experiment published~\cite{BaBar02} similar
results based on 20.3~fb$^{-1}$ data, showing clear excess beyond
NRQCD expectations~\cite{Beneke}, of order $10^{-3}$ in rate, for
$p^*_{J/\psi}\lesssim 0.8$ GeV.

As the excess involves slow moving $J/\psi$ mesons, it must have
hadronic, rather than perturbative, origins.
Various proposals have been advanced.
The suggestion of $B\to J/\psi\bar\Lambda p$~\cite{Brodsky} has
been studied recently by BaBar~\cite{BaBar03}; the event rate at
$10^{-5}$ order cannot explain the slow $J/\psi$ excess.
Intrinsic charm content of the $B$ meson could lead to $B\to
J/\psi D(+\pi)$ final states~\cite{IC}, which can in principle
explain the data, but experimental studies are not yet
forthcoming. If $B\to J/\psi D\pi$ dominates, the slow pion does
not pair with $D$ to form a $D^*$, and would pose a challenge.
Another possibility~\cite{Eilam} would be $B\to J/\psi K_g$, where
$K_g$ is a hybrid meson with $\bar sgq$ constituents. A recent
estimate~\cite{Close} suggests that the rate could be in the
ballpark.

In this note we take a heuristic approach to explore the last
possibility. We find the hybrid scenario is indeed viable. We
suggest the signature of $B\to J/\psi + K+n\pi$, where $n$ cannot
be more than a few, should be experimentally searched for. If the
$K+n\pi$ system tends to peak at some mass, but does not descend
from $D$ meson decay, then the hybrid meson picture could be
substantiated.

Let us visualize why a hybrid meson recoiling against a $J/\psi$
could be the right picture. In $b\to c\bar c s$ decay, the $c\bar
c$ pair is dominantly formed in a color octet configuration, hence
charmonium production is color-suppressed. Imagine that, upon $b$
quark weak decay, the $c$ and $\bar c$ quarks are moving apart
with more than $\sim 1$~GeV kinetic energy. Soft ``muck" effects
cannot change the configuration, and the system would tend to
break up into open charm meson pairs, resulting in $D^{(*)}\bar
D_s^{(*)}$ or $D\bar D\bar K$ final states. But if the $c$ and
$\bar c$ momenta are relatively colinear, it can be viewed as a
``proto-charmonium". Because of the heaviness of $m_c$, this
small, dominantly color octet $c\bar c$ system would recoil
against the $s$ quark, again relatively unperturbed by the soft
``muck". By the time it separates from the $s$ quark by order 1
fm, strong, nonperturbative effects set in: it has to hadronize.
But since this is already a ``proto-charmonium", i.e. the spatial
and spin wave-function already maps well onto some physical
charmonium state, the only problem is it has to shed color. An
effective color octet charge is thus left to neutralize the $s\bar
q$ system, and the simplest configuration is that of $sg\bar q$,
which we call an $\bar K_g$ hybrid system.

Whether a $K_g$ hybrid meson really exists becomes semantical. As
visualized above, the leftover color octet charge with the color
octet $s\bar q$ system is not by perturbative gluon
emission~\cite{Eilam}, but by the fact that two separate color
strings extend from the $s$ and the $\bar q$ towards the point
where the $c\bar c$ hadronization took place. It is plausible that
this $s$--$g$--$\bar q$ string configuration could resonate and
the amplitude gets enhanced, and energy-momentum is exchanged with
the departing charmonium. If the $K_g$ resonance is of order
2~GeV~\cite{Isgur} in mass, then the $J/\psi$ has to be slow. We
have therefore constructed the physical picture whereby slow
$J/\psi$ (charmonium in general) can receive enhancement.

Let us illustrate further this picture. The relevant effective
weak Hamiltonian is
\begin{eqnarray}
H_{\rm W}&=&{G_F\over\sqrt2} V^*_{cs} V_{cb}
          (c_1\,{\cal O}_1+c_2\,{\cal O}_2),
\nonumber\\
{\cal O}_{1(2)}&\equiv&\overline s_\alpha
\gamma^\mu(1-\gamma_5)c^{\alpha(\beta)}\,
          \overline c_{\beta(\alpha)} \gamma_\mu(1-\gamma_5)b^\beta.
\end{eqnarray}
The amplitude for $\bar B\to J/\psi \bar K$ decay is
\begin{equation}
A(J/\psi \bar K)=\frac{G_{\rm F}}{\sqrt2} V^*_{cs} V_{cb} \, a_2
\, \langle J/\psi\vert \bar c \gamma^\mu c \vert 0\rangle \langle
\bar K\vert \bar s \gamma_\mu b\vert \bar B\rangle,
\label{eq:K}
\end{equation}
where $a_2$ is the effective coefficient governing
color-suppressed modes, which in the naive factorization limit is
$c_2+c_1 /3$.
Fitting data gives $a_2\gtrsim 0.25$, but factorization
calculations tend to give lower values of $a_2\lesssim$ 0.2,
suggesting that nonfactorized effects are important.

We now infer from Eq. (\ref{eq:K}) by analogy the formula for
$B\to J/\psi K_g$.
It is well known that
\begin{equation}
{\cal O}_1=2\,(\overline c\,T^a c)_{V-A} (\overline s\,T^a
b)_{V-A}+\frac{{\cal O}_2}{3} ,
\end{equation}
where $T^a$ is the color SU(3) generator. A color octet $\bar c c$
pair is favored.
The ``proto-$J/\psi$" would still be produced by a $\bar cc$
vector current, but stripping off the extra color as it departs,
one is left with a ``constituent" gluon in association with the
$s\bar q$ bilinear.
The nonperturbative picture should be two (different colored)
strings extending from the point of departure of the $J/\psi$
towards the recoiling $s$ and spectator $\bar q$ quarks.
Using a factorization language for sake of illustration, the
matrix element product, $\langle J/\psi\,g|(\bar c T^a
c)_{V-A}|0\rangle \; \langle (s\bar q)_8|(\bar s\,T^a
b)_{V-A}|\bar B\rangle$, where $g$ is a constituent gluon and $(s
\bar q)_8$ is a color octet quark pair, should be nonvanishing.
The final state therefore has the $J/\psi$ recoiling against a
color singlet $s \bar q g$ configuration, which could form a
hybrid $K_g$ meson. We heuristically write the operator $2\,(\bar
c\,T^a c)_{V-A} (\bar s\,T^a b)_{V-A}$ as $\bar c\gamma^\mu
c\,[\bar sgb]_\mu$, which we now take as our Ansatz.
%
%
%
%
%
%
Thus, the formula for $\bar B\to J/\psi \bar K_g$ becomes,
\begin{equation}
A(J/\psi \bar K_g) \propto \frac{G_{\rm F}}{\sqrt2} V^*_{cs}
V_{cb} \, c_1 \, \langle J/\psi\vert \bar c \gamma^\mu c \vert
0\rangle \langle \bar K_g\vert [\bar s gb]_\mu \vert \bar
B\rangle,
 \label{eq:Kg}
\end{equation}
where the $[\bar sgb]$ operator can convert the $\bar B$ meson
into $\bar K_g$, including matching its $J^P$, in analogy with the
$\bar s\gamma_\mu b$ current converting $\bar B$ into $\bar
K^{(*)}$.
%
%
Note that the decay amplitude is proportional to $c_1\sim 1$
rather than $a_2$, but there is some proportionality constant,
{\it hopefully of order 1}, from the Ansatz
we made above.
Due to the usual difficulty of hadronic physics and the model
dependence that would necessarily arise, we do not attempt at
calculating theoretically this proportionality constant, but turn
to data for its determination.

The simplest situation would be to have a $K_g$ meson with $J^P =
0^{\mp}$ (note that there are no ``exotic" kaon hybrids). The
contraction of vector current indices should then be very similar
between Eqs. (\ref{eq:K}) and (\ref{eq:Kg}), and the $B\to J/\psi
K_g(0^{\mp})$ decay rate is estimated to be
\begin{equation}
\kappa_0^2 \left\vert\frac{c_1}{a_2}\right\vert^2 \frac
{p^3_{K_g}}{p^3_{K}} |{\rm BW}|^2 \, {\mathcal B}(B\to J/\psi K),
 \label{eq:KgovK}
\end{equation}
where $\kappa_0$ is related to the aforementioned proportionality
factor, but it also contains possible differences in $B\to K_g$
and $B\to K$ form factors, and $p_{K_{(g)}}$ is the momentum of
$K_{(g)}$ in the $B$  decay frame. Since the hybrid $K_g$ meson is
expected to be broad, the decay rate is modulated by the
Breit-Wigner factor~\cite{PDG}
\begin{equation}
{\rm BW}(q^2)=\frac{\sqrt{q^2}\,\Gamma(q^2) }{(q^2-m^2)+i
\sqrt{q^2} \Gamma(q^2)},
 \label{eq:BW}
\end{equation}
where $q^2=m_{J/\psi}^2+m_B^2-2 m_B E_{J/\psi}$ and $E_{J/\psi}$
is the $J/\psi$ energy in the $B$ rest frame.
Note that, to account for kinematic dependence~\cite{PDG}, a
$\sqrt{q^2}$ factor is used instead of $m$. Furthermore,
\begin{equation}
\Gamma(q^2)=\sqrt{q^2} \, \frac{\Gamma_0}{m},
 \label{eq:width}
\end{equation}
where $m$, $\Gamma_0$ are the mass and width of $K_g$ at
$q^2=m^2$.

We now make a fit to the inclusive $J/\psi$ spectrum and see
whether a single hybrid $K_g$ suffices to account for the observed
excess.
The direct (feeddown subtracted) $B\to J/\psi X$ data is taken
from Ref.~\cite{BaBar02}. BaBar uses NRQCD plus $B\to J/\psi
K^{(*)}$ simulation results to fit their data. To simplify, we
follow Refs.~\cite{Brodsky,IC}, and use
\begin{equation}
f(p)=N (p-p_{\rm min})(p-p_{\rm max}) \exp\left[-\frac{(p-\bar
p)^2}{\sigma^2_0}\right],
\end{equation}
to mimic the color-octet NRQCD~\cite{Beneke} and $J/\psi
K^{(*)}$~\cite{BaBar02} components.
We find $(N,\ p_{\rm min},\ p_{\rm max},\ \bar p,\ \sigma_0) =
(26,\ 0,\ 1.95,\ 1.21,\ 0.5)$ and $(180,\ 1.2,\ 1.95,\ 1.65,\
0.3)$, with all energy-momentum in GeV, give good accounts of the
two contributions, respectively.

\begin{figure}[t!]
\centerline{
            {\epsfxsize3 in \epsffile{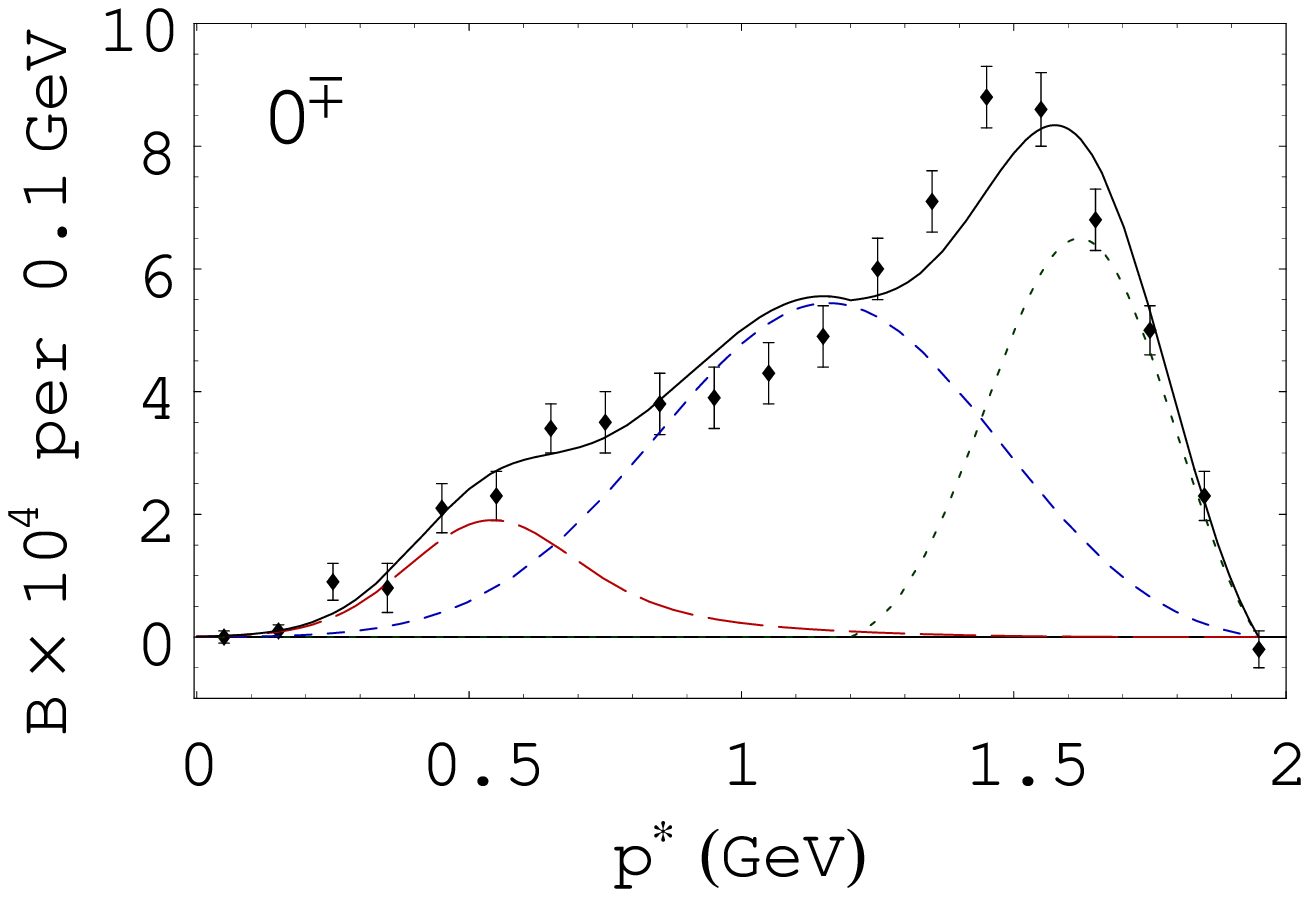}}
}
\centerline{
            {\epsfxsize3 in \epsffile{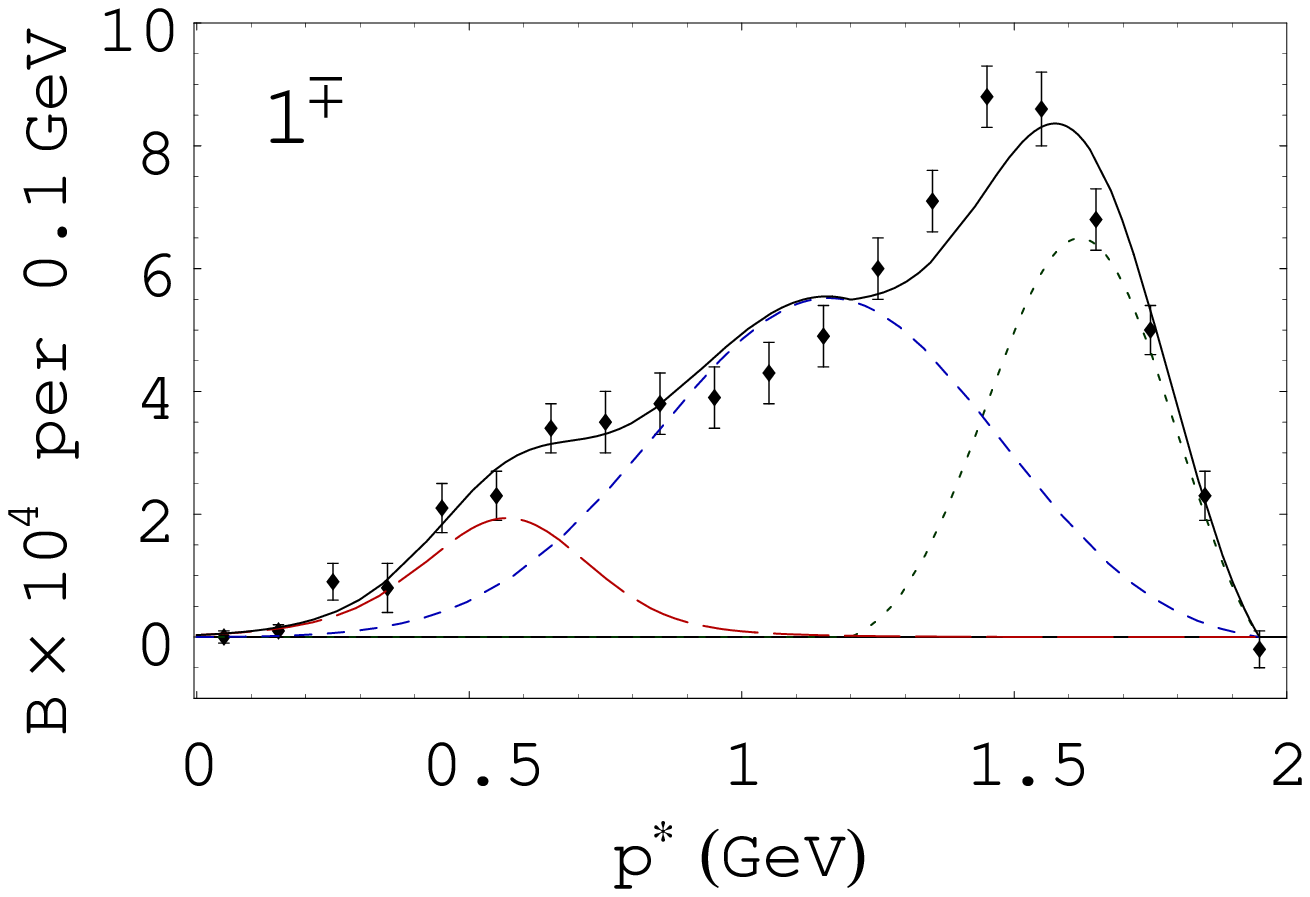}}
}
\caption{
 $B\to J/\psi+X$ decay spectrum. Data is from Ref.~\cite{BaBar02}.
The dotted and short-dashed lines correspond to $J/\psi K^{(*)}$
simulation~\cite{BaBar02} and NRQCD component~\cite{Beneke}. The
long-dashed line corresponds to the $B\to J/\psi K_g$ contribution
for (a) $0^\mp$ and (b) $1^\mp$ case, with fitted mass and width
given in text.}
 \label{fig:spectrum}
\end{figure}

Adding now the $B\to J/\psi K_g$ contribution of
Eq.~(\ref{eq:KgovK}), 
we smear by a Gaussian with rms spread of 0.12~GeV~\cite{BaBar02}
to account for broadening in the $\Upsilon(4S)$ frame. Using
$|c_1/a_2|=5$ for illustration, and the isospin averaged
${\mathcal B}(B\to J/\psi K)=0.94\times 10^{-3}$, and allowing the
NRQCD contribution to float in the new fit, we obtain
\begin{equation}
\kappa_0\simeq 2.3,\ m\simeq 2.08\;{\rm GeV},\ \Gamma_0\simeq
72\;{\rm MeV}.
 \label{eq:fit}
\end{equation}
The NRQCD contribution is reduced by 12\% with respect to
Ref.~\cite{BaBar02}, the fitted $B\to J/\psi K_g$ branching ratio
is $8.5\times 10^{-4}$, and the spectrum is shown in
Fig.~\ref{fig:spectrum}(a).
It is remarkable that a single $0^\mp$ hybrid $K_g$ meson with
mass $\sim$ 2.1 GeV and width $\sim$ 100 MeV could account for the
observed slow $J/\psi$ excess. The fudge factor $\kappa_0\sim$ 2.3
means our inference by analogy is in the right ballpark.

Let us investigate the case for $K_g$ meson with $J^P = 1^\mp$.
Replacing $K$ by $K^*$ in Eq. (\ref{eq:K}), we estimate, in
analogy to Eq. (\ref{eq:KgovK}), the $B\to J/\psi K_g(1^{\mp})$
decay rate to be
\begin{equation}
\kappa_1^2 \left\vert\frac{c_1}{a_2}\right\vert^2 \frac
{p_{K_g}}{p_{K^*}} |{\rm BW}|^2 \,{\mathcal B}(B\to J/\psi K^*),
 \label{eq:Kg1ovKst}
\end{equation}
where $\kappa_1$ is analogous to $\kappa_0$, but now the $B\to
K_g$ and $B\to K^*$ form factor ratio can be rather complicated,
because of two possible helicity configurations. Absorbing all of
this into $\kappa_1$, we retain linear momentum dependence
corresponding to longitudinally polarized component, which is
expected to be dominant from usual form factor models, as well as
in the heavy quark limit.

Performing a fit as before using isospin averaged ${\mathcal
B}(B\to J/\psi K^*)=1.35\times 10^{-3}$, we obtain
\begin{equation}
\kappa_1\simeq 0.6,\ m\simeq 2.05\;{\rm GeV},\ \Gamma_0\simeq
70\;{\rm MeV}.
 \label{eq:fit1}
\end{equation}
The NRQCD contribution is reduced by  11\% with respect to
Ref.~\cite{BaBar02}, and the fitted ${\mathcal B}(B\to J/\psi
K_g(1^\mp))$ is $7.9\times 10^{-4}$. The spectrum is shown in
Fig.~\ref{fig:spectrum}(b), which is similar to
Fig.~\ref{fig:spectrum}(a) since mass and width are almost the
same.
Note that the fudge factor $\kappa_1\simeq 0.6$ appears even more
reasonable than the $0^\mp$ case.

\begin{figure}[t!]
\centerline{
            {\epsfxsize3 in \epsffile{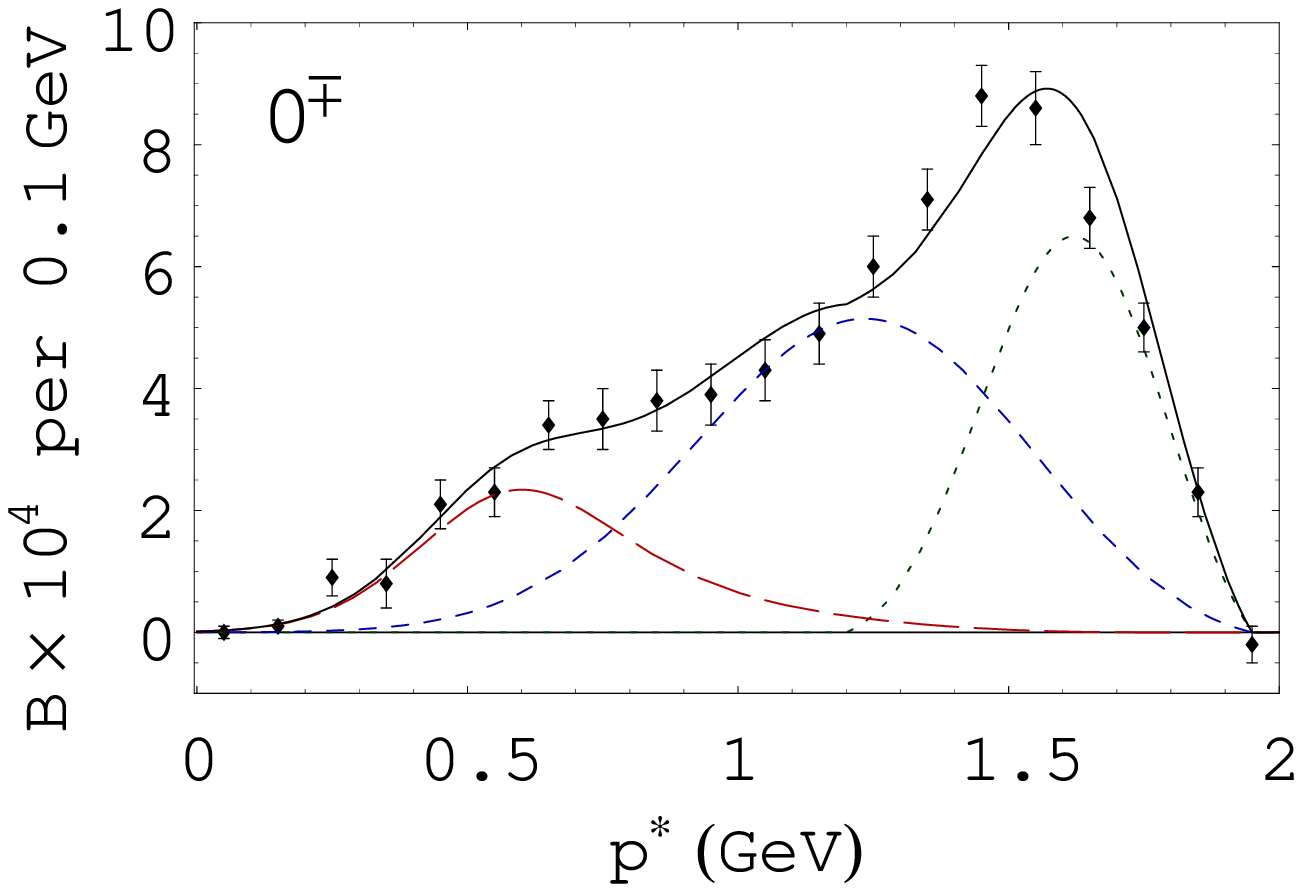}}
}
\centerline{
            {\epsfxsize3 in \epsffile{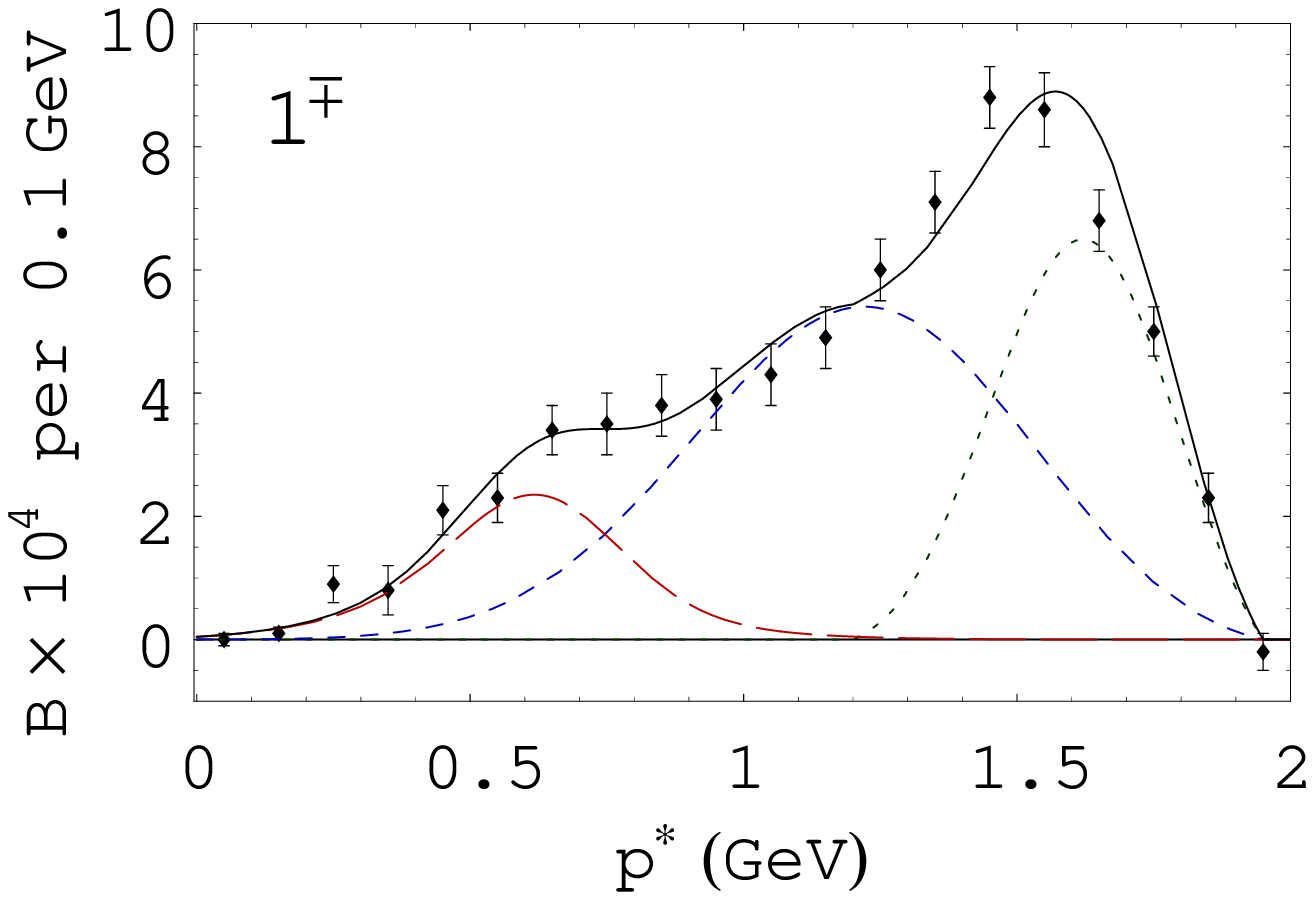}}
}
\caption{
 Same as Fig. 1 but allowing the NRQCD
component to float for an improved fit.}
 \label{fig:spectrum2}
\end{figure}

Fig. \ref{fig:spectrum} suggests that the fits may not yet be
optimized for $p^*_{J/\psi}$ between 0.9--1.5 GeV. As we have
allowed some freedom in the strength of the NRQCD
contribution~\cite{Beneke}, we now allow its peak position to
float as well. Fitting again, we find
for $0^\mp$ case
\begin{equation}
\kappa_0\simeq 2.0,\ m\simeq 2.08\;{\rm GeV},\ \Gamma_0\simeq
147\;{\rm MeV},
 \label{eq:fitp}
\end{equation}
with the NRQCD contribution reduced by 13\% with respect to
Ref.~\cite{BaBar02}, and the parameter $\bar p$ shifted by 100~MeV
to 1.31~GeV. The fitted $B\to J/\psi K_g$ branching ratio becomes
$12.9\times 10^{-4}$.
For $1^\mp$ case, we find
\begin{equation}
\kappa_1\simeq 0.5,\ m\simeq 2.03\;{\rm GeV},\ \Gamma_0 \simeq
103\;{\rm MeV},
 \label{eq:fit1p}
\end{equation}
with NRQCD contribution reduced by 10\% with respect to
Ref.~\cite{BaBar02}, $\bar p$ shifted from 1.21 GeV to 1.29~GeV,
and fitted ${\mathcal B}(B\to J/\psi K_g(1^\mp)) \simeq 10.5\times
10^{-4}$.
The fitted spectrum is shown in Fig.~\ref{fig:spectrum2}. The
remaining slight ``discrepancy" can be attributed to the
difference between a perturbative vs. hadronic approach, e.g.
summing over $J/\psi K_1$~\cite{psiK1}, $J/\psi K_2^*$, etc.
modes.

We do not commit ourselves to what should be the lightest $K_g$
hybrid state, or how large is the splitting for further
excitations. It is gratifying, however, that the fitted masses of
order 2--2.1 GeV is close to expectations~\cite{Isgur}. The width
of 70--150 MeV may seem narrow, but kaonic hybrids have not been
widely discussed in the literature, and the relative narrowness
would make experimental identification easier.
As for decay modes, we remark that flux tube models
suggest~\cite{Isgur2} hybrids decay into final states with one
excited meson. Thus, one should consider reconstructing $K_g$ in
$K_{0,1,2}^*\pi(\rho)$, $K^{(*)} f_{0,1,2}$, maybe also $K
f_2^\prime$ final states.
It would be fascinating if a heavy kaon resonance is found to be
dominating the slow $J/\psi$ excess from $B$ decay.

We note that the $K_g$ width is larger for the improved fit of
Fig.~2. We have checked that $\Gamma_0\sim$ 250~MeV is possible,
if the ``peak position" parameter $\bar p$ is allowed to shift
slightly higher, to 1.34 GeV and 1.31 GeV, respectively, for the
$0^\mp$ and $1^\mp$ cases. Thus, the narrowness of $\Gamma_0 \sim
70$ MeV of Eqs.~(\ref{eq:fit}) and (\ref{eq:fit1}) may be an
artefact of trying to mimic the NRQCD result of
Ref.~\cite{Beneke}. The latter work subtracted $B\to J/\psi K$ and
$J/\psi K^*$, an approach BaBar adopted, but it was done before
the Belle measurement~\cite{psiK1} of ${\mathcal B}(B\to J/\psi
K_1(1270)) \simeq 1.55\times 10^{-3}$ (isospin averaged), which is
comparable to $B\to J/\psi K$ and $J/\psi K^*$. An update of
Ref.~\cite{Beneke} would be helpful.

The $0^-$ and $1^-$ hybrid quantum numbers allow us to make some
insight into the possible cause of sizable nonfactorizable
contributions to $B\to J/\psi K^{(*)}$, $J/\psi K_1$ decay. The
physical $K^{(*)}$ state may have a hybrid Fock component,
$\vert K\rangle = c_\theta \vert K^{(0)}\rangle + s_\theta \vert
K_g^{(0)}\rangle$.
One then finds
\begin{equation}
a_2^{\rm eff.} = a_2^{\rm fac.} \left( c_\theta + \frac{c_1}{a_2}
\kappa s_\theta \right),
 \label{eq:a2eff}
\end{equation}
where $a_2^{\rm fac.}$ is from factorization calculations,
typically of order 0.15--0.2. One sees from our fitted $\kappa$
values that a hybrid admixture of a few \% to no more than 10\% in
the $K^{(*)}$ wavefunction can suffice to account for the large
$a_2^{\rm eff.} \sim$ 0.25--0.3 extracted from data, in large part
because of gaining the $c_1/a_2$ factor.

%

We comment on a recent proposal that slow $J/\psi$ from $B$ decay
could arise from $cq\bar q\bar c$ four quark states~\cite{cqqq}.
The physical picture would be that the color octet $c\bar c$ picks
up a color octet $q\bar q$ pair as it hadronizes. Ref.~\cite{cqqq}
gave some arguments for why $J/\psi$ may end up slow.
It may not be so easy to distinguish this proposal from the
present one, as both lead to $J/\psi + K + n\pi$ final states, and
in fact both mechanisms may well be at work concurrently.
To distinguish the two mechanisms, one would have to check whether
the charmonium system is resonating with some of the recoil
hadrons. For $K_g$ mechanism, it would be helpful if
$K_{0,1,2}^*\pi(\rho)$, $K^{(*)} f_{0,1,2}$, $K f_2^\prime$
dominance in final state is borne out.
We remark that for most pictures, one should also find excess in
slow $\eta_c$ mesons.

In summary, we have illustrated that a single hybrid $K_g$ state
recoiling against a $J/\psi$ could explain the slow $J/\psi$
excess observed in $B$ decay with rate at $10^{-3}$ order. The
fitted $K_g$ mass is of order 2.05--2.1 GeV, with width of order
70--150 MeV, but could be broader. Experimental signature would be
to reconstruct $K_g\to K+n\pi$, probably in
$K_{0,1,2}^*\pi(\rho)$, $K^{(*)} f_{0,1,2}$, $K f_2^\prime$
configurations.

\vskip 0.3cm
\noindent  
We thank Hsuan-Cheng Huang for discussions, and Jim Mueller for
pointing out Ref.~\cite{Close} to us. This work is supported in
part by NSC grants 91-2112-M-002-027, 91-2811-M-002-043, the MOE
CosPA Project, and the BCP Topical Program of NCTS.

\end{document}